\documentclass[pre, onecolumn]{revtex4}
\usepackage{amsmath}
\usepackage{amsfonts}
\usepackage{lipsum}
\usepackage{braket}
\usepackage{bm}
\usepackage[dvipdfmx]{graphicx}
\usepackage{xcolor}
\usepackage [] {enumitem}

\usepackage{ulem}

\begin{document}
\title{Floquet-Gibbs state in open quantum systems}
\author{Tatsuhiko Shirai$^1$}
\email{shirai@exa.phys.s.u-tokyo.ac.jp}
\author{Takashi Mori$^2$}
\author{Seiji Miyashita$^2$}
\affiliation{%
$^1$The Institute for Solid State Physics, University of Tokyo, 5-1-5 Kashiwanoha, Kashiwa, Chiba 277-8581, Japan\\
}
\affiliation{%
$^2$Department of Physics, Graduate School of Science,
The University of Tokyo, 7-3-1 Hongo, Bunkyo-Ku, Tokyo 113-8656, Japan\\
}

\begin{abstract}
We study long-time asymptotic states of periodically driven quantum systems coupled to a thermal bath. In order to describe a class of such a system, we introduce the Floquet-Gibbs state, i.e. the state whose density matrix is diagonalized in the basis of the Floquet state of the system Hamiltonian, and its diagonal element obeys the Boltzmann distribution over its Floquet quasienergy. We obtain sufficient conditions for the realization of the Floquet-Gibbs state in a system with infinitesimal system-bath coupling [T. Shirai, et al., Phys. Rev. E {\bf 91}, 030101 (2015)]. These conditions severely restrict a class of suitable physical models attaining the Floquet-Gibbs state. We also show that some of the conditions can be lifted by imposing conditions on timescales of the thermal bath with the aid of the truncated Floquet Hamiltonian in the Floquet-Magnus expansion [T. Shirai, et al., New Journal of Physics {\bf 18}, 053008 (2016)]. In this paper we give a overview of this theory and reconsider it by looking at the dynamics from a rotating frame.
\end{abstract}

\maketitle

\section{Introduction}
When a quantum system is in contact with a thermal bath,
which consists of infinite number of degrees of the freedom,
relaxation dynamics emerges in the reduced density matrix of the system.
When the system is time independent and the system-bath coupling is infinitesimal,
the long-time asymptotic state is described by the Gibbs state, i.e.
\begin{equation}
\rho_{\beta}^{\rm can} = \frac{e^{-\beta H}}{{\rm Tr}e^{-\beta H}},
\end{equation}
where $H$ is the Hamiltonian of the system of interest and $\beta$ is the inverse temperature of the thermal bath.
The simple form allows us to investigate thermal properties such as the specific heat or magnetization without reference to the complicated relaxation dynamics.

Here we extend this simple expression of the Gibbs form to quantum systems with a periodically driving field,
whose Hamiltonian is given by
\begin{equation}
H(t)=H(t+T),
\end{equation}
where $T$ is the period of the driving field.
This type of the Hamiltonian displays one of the typical non-equilibrium situations,
and is implemented in wide range of fields such as quantum optics, condensed matter physics, and cold atom physics.
Due to the time periodicity of the Hamiltonian, we can introduce a time-independent Hamiltonian called the Floquet Hamiltonian $H_{\rm F}$, which is defined by a time-evolution operator over one period~\cite{bukov2015universal,eckardt2015high} as
\begin{equation}
e^{-\frac{i}{\hbar} H_{\rm F} T} \equiv {\cal T}e^{-\frac{i}{\hbar} \int_0^T H(t) dt},
\end{equation}
where ${\cal T}$ is a usual time-ordering operator.
When we look at the coherent dynamics stroboscopically at $t=nT$ $(n \in \mathbb{Z})$, the state is obtained by simply multiplying $\exp (-i H_{\rm F} t/\hbar)$ on the initial state at $t=0$.
In this sense the Floquet Hamiltonian $H_{\rm F}$ plays a role of $H$ in dynamics, and thus it is relevant for the description of the isolated periodically driven systems.
Here we investigate the {\it thermodynamic} relevance of the Floquet Hamiltonian.
Namely we introduce the Floquet-Gibbs state,
\begin{equation}
\rho_{\rm FG} \equiv \frac{e^{-\beta H_{\rm F}}}{{\rm Tr} e^{-\beta H_{\rm F}}}.
\end{equation}
and discuss whether or not the driven system in contact with a thermal bath is relaxed to this state after a sufficiently long time.

The general asymptotic states of the periodically driven systems are extremely complex, and hence we restrict ourselves to a class of physical models.
We suppose that the system is excited by a strong and high-frequency driving field.
To express it explicitly, we divide the system Hamiltonian into two parts,
\begin{equation}
H(t)=H_0 + \Omega H_{\rm ex}(\Omega t),
\end{equation}
where $H_0$ and $\Omega H_{\rm ex} (\Omega t)$ represent a static part of the Hamiltonian and the driven part of the Hamiltonian, respectively.
The frequency of the driving field is denoted by $\Omega$, i.e. $\Omega = 2\pi/T$.
The amplitude is taken to be proportional to the frequency so that the effects of the driving field appear in dynamics even at high frequency.
It is noted here that we do not take the limit of $\Omega \to \infty$, in which the time-dependent Hamiltonian is reduced to a time-independent one.
We consider high but finite frequency.
The theoretical framework for the relaxation dynamics is formulated by considering a thermal bath in addition to the system of interest.
Namely the Hamiltonian of the total system reads
\begin{equation}
H_{\rm T}(t)=H(t)+H_{\rm B} +H_{\rm SB},
\end{equation}
where $H_{\rm B}$ is a bath Hamiltonian and $H_{\rm SB}$ is an interaction Hamiltonian between the system of interest and the thermal bath.

Here we discuss the thermodynamic relevance of $H_{\rm F}$ in a rotating frame,
in which the wave function in the rotating frame $\ket{\psi_{\rm R}(t)}$
is related to that in the static frame $\ket{\psi (t)}$ by
\begin{equation}
\ket{\psi(t)} ={\cal T}e^{-\frac{i}{\hbar}\int_0^t H_{\rm ex}(\tau)d\tau} \ket{\psi_{\rm R}(t)} \equiv U(t)\ket{\psi_{\rm R}(t)}.
\end{equation}
The dynamics in the rotating frame is governed by the following Hamiltonian,
\begin{equation}
H_{\rm T, R}(t)=H_{\rm R}(t)+H_{\rm B}+U^{\dagger}(t) H_{\rm SB} U(t),
\end{equation}
where
\begin{equation}
H_{\rm R}(t)=U^{\dagger}(t) \left( H(t)-i \hbar \frac{\partial}{\partial t} \right) U(t)=U^{\dagger}(t) H_0 U(t).
\end{equation}
In the rotating frame the effects of the driving field appear only through the unitary operator $U(t)$.
When the time dependences, i.e. $H_{\rm R}(t)$ and $U^{\dagger}(t) H_{\rm SB} U(t)$, are effectively eliminated,
the Floquet-Gibbs state is expected to appear.
We will discuss in what conditions this time dependence can be eliminated for each term separately.

The relaxation dynamics of the system of interest with a weak system-bath coupling is described by a Markovian quantum master equation:
\begin{equation}
\frac{d\rho(t)}{dt}=-\frac{i}{\hbar} [H_{\rm R}(t), \rho(t)]+\gamma \Gamma[\rho(t)],\label{relaxation_dynamics}
\end{equation}
where $\rho (t)$ is the reduced density matrix of the system of interest in the rotating frame.
The first term shows a coherent dynamics under the system Hamiltonian $H_{\rm R}(t)$
and the second term represents the dissipation effects due to the coupling to the thermal bath.
The strength of the system-bath coupling is denoted by $\gamma$ and the form of the dissipation operator is simply expressed by $\Gamma [\cdot]$.
After a sufficiently long time, the reduced density matrix approaches a time-periodic state,
and thus if we look at the dynamics stroboscopically it converges as
\begin{equation}
\rho_{\rm asy}=\lim_{n \to \infty} \rho(nT), \label{asymptotic}
\end{equation}
which is here compared with the Floquet-Gibbs state.

In this paper, we first discuss quantum systems with infinitesimal system-bath coupling $\gamma \to 0$.
It is noted that its dynamics is different from that of the isolated system because we take simultaneously observation time $t$ to be infinite keeping $\gamma t$ fixed.
Next we discuss the $\gamma$ dependence.
The coupling strength $\gamma$ determines the timescale for the relaxation dynamics,
which can affect the long-time asymptotic state significantly.
For the theoretical framework of quantum master equations, see~\cite{kohn2001periodic,breuer2000quasistationary,breuer2002theory,ketzmerick2010statistical,langemeyer2014energy,shirai2015condition} for systems with infinitesimal system-bath coupling $\gamma \to 0$ and~\cite{blumel1991dynamical,gasparinetti2013environment,shirai2016effective} for systems with weak but finite system-bath coupling.

\section{Floquet-Gibbs state for systems with infinitesimal system-bath coupling}
For quantum systems with infinitesimal system-bath coupling,
we obtain sufficient conditions for the realization of the Floquet-Gibbs state:
\begin{enumerate}[label={\arabic*.}]
\item The frequency of the driving field is much larger than the spectral width of $H_0$, $\hbar \Omega \gg \| H_0 \|$,
\item The driving Hamiltonian commutes with itself at different instants of time,\\
$[H_{\rm ex}(t_1), H_{\rm ex}(t_2) ]=0$,
\item The Hamiltonian for the system-bath coupling commutes with the driving Hamiltonian, $[H_{\rm SB}, H_{\rm ex}(t)] =0$,
\end{enumerate}
where $\| \cdot \|$ denotes the operator norm.
The ergodic property in the high-frequency limit is also assumed.
In the following we will explain the physical significances of these conditions.

The first and second conditions are necessary to replace $H_{\rm R}(t)$ by $H_{\rm F}$.
In order to explain this, we consider a one-half spin system whose Hamiltonian is given by
\begin{equation}
H(t)=h^x \sigma^x + \hbar \Omega \cos (\Omega t) \sigma^z,
\end{equation}
where $\{\sigma^{\alpha}\}_{\alpha = \{ x, y, z\}}$ are the Pauli matrices.
The first term represents the Zeeman energy due to a static magnetic field along the $x$-axis, and the second term describes a strong and high-frequency driving field along the $z$-axis.
The first term and the second term corresponds to $H_0$ and $\Omega H_{\rm ex}(\Omega t)$, respectively.
It is noted that the driving Hamiltonian satisfies the second condition, i.e. $[H_{\rm ex}(t_1), H_{\rm ex}(t_2)]=0$.
We then obtain a simple form of $U(t)$ as
\begin{equation}
U(t)\equiv{\cal T} e^{- \frac{i}{\hbar} \int_0^t \Omega H_{\rm ex}(\tau) d\tau} = e^{- \frac{i}{\hbar} \int_0^t \Omega H_{\rm ex}(\tau) d\tau} =e^{-i \sin(\Omega t)\sigma^z}.\label{cond2}
\end{equation}
Here we have used the second condition to remove the time-ordering operator.
The unitary operator is time periodic and oscillates at high frequency.
The system Hamiltonian in the rotating frame also oscillates at high frequency,
which is explicitly expressed by
\begin{equation}
H_{\rm R}(t) =h^x \left[ \sigma^x \cos \left( \sin \Omega t \right) -\sigma^y \sin \left( \sin \Omega t \right) \right].
\end{equation}
In spite of the large oscillating amplitude in the static frame, in the rotating frame it is the order of $h^x$, which is not so strong due to the first condition, $\hbar \Omega \gg \| H_0 \|$.
Since in the rotating frame not strong but high-frequency driving field is applied, the dynamics is well approximated by the time average of the rotating Hamiltonian,
which is time independent and close to the Floquet Hamiltonian up to the leading order of $\Omega^{-1}$ (see Eq.~(\ref{Magnus_eq})),
\begin{equation}
H_{\rm R}(t) \simeq \frac{1}{T}\int_0^T H_{\rm R}(t) dt \simeq H_{\rm F}.\label{time_average}
\end{equation}

The third condition, $[H_{\rm SB}, H_{\rm ex}(t)]=0$, is necessary to eliminate the time dependence of the interaction Hamiltonian in the rotating frame:
\begin{equation}
U^{\dagger}(t) H_{\rm SB} U(t) =H_{\rm SB}.
\end{equation}

As a result when all the three conditions are satisfied, the total Hamiltonian in the rotating frame is approximately given by
\begin{equation}
H_{\rm R}(t)+H_{\rm B}+ U^{\dagger}(t) H_{\rm SB} U(t) \simeq H_{\rm F}+H_{\rm B}+H_{\rm SB}.
\end{equation}
This system is nothing but a time-independent system in contact with a thermal bath, and therefore the system is expected to approach to the Floquet-Gibbs state.
To explicitly show this fact we suppose that the system-bath coupling is infinitesimal, i.e. $\gamma \to 0$, in which the relaxation dynamics is described by the Lindblad equation.
Setting the interaction Hamiltonian as
\begin{equation}
H_{\rm SB}=X \otimes Y,
\end{equation}
where $X$ and $Y$ are the operator of the system and the thermal bath, respectively, the diagonal elements of the density matrix in the eigenbasis of $H_{\rm F}$, i.e. $H_{\rm F}\ket{\phi_p}=\epsilon_p \ket{\phi_p}$, obey
\begin{equation}
\frac{d\rho_{pp}(t)}{dt}=\sum_{q} |\bra{\phi_p}X \ket{\phi_q}|^2 (G(\omega_{pq})\rho_{qq}(t) - G(\omega_{qp}) \rho_{pp}(t)),\label{Lindblad}
\end{equation}
where $\rho_{pp}(t) \equiv \bra{\phi_p} \rho(t) \ket{\phi_p}$ and $\omega_{pq}=(\epsilon_p -\epsilon_q)/\hbar$.
The effects of the thermal bath are taken into account in the correlation function of the thermal bath as
\begin{equation}
G(\omega)=\int_{-\infty}^{\infty} {\rm Tr} \left[ e^{iH_{\rm B}t} Y e^{-i H_{\rm B}t} Y \frac{e^{-\beta H_{\rm B}}}{{\rm Tr} e^{-\beta H_{\rm B}}} \right] e^{-i \omega t} dt,
\end{equation}
which satisfies the Kubo-Martin-Schwinger (KMS) relation~\cite{kubo1957statistical},
\begin{equation}
G(-\omega)=G(\omega) e^{\beta \hbar \omega}.
\end{equation}
The equations for the diagonal elements obey the Pauli master equation or the rate equation, and thus its dynamics is characterized by the transition probability from $\ket{\phi_p}$ to $\ket{\phi_q}$:
\begin{equation}
T_{p \to q}= |\bra{\phi_p} X \ket{\phi_q}|^2 G(\omega_{qp}).\label{transition}
\end{equation}
It is easily confirmed that the detailed balance condition is satisfied using the KMS condition;
\begin{equation}
\frac{T_{p \to q}}{T_{q \to p}}=\frac{G(\omega_{qp})}{G(\omega_{pq})}=e^{\beta(\epsilon_p - \epsilon_q)}.\label{detailed}
\end{equation}
Thus, as far as the ergodicity is satisfied, the diagonal element approaches the Boltzmann distribution over its Floquet quasienergy;
\begin{equation}
\lim_{t\to \infty}\rho_{pp}(t)= \frac{e^{-\beta \epsilon_p}}{\sum_q e^{-\beta \epsilon_q}}.\label{Boltzmann}
\end{equation}
The off-diagonal elements, i.e. $\rho_{pq}(t) \equiv \bra{\phi_p} \rho(t) \ket{\phi_q}$ for $p \neq q$, obey
\begin{equation}
\frac{d\rho_{pq}(t)}{dt}=\left[ -i(\tilde{\epsilon}_p-\tilde{\epsilon}_q) -\frac{\Gamma_{pq}}{2} \right] \rho_{pq}(t).
\end{equation}
\begin{equation}
\left\{
\begin{aligned}
\tilde{\epsilon}_p=&\epsilon_p+\sum_r |\bra{\phi_r} X \ket{\phi_p}|^2 \int_{-\infty}^{\infty} \frac{d\omega'}{2\pi} \frac{{\cal P}}{\omega'-\omega_{rp}} G(\omega'), \nonumber\\
\Gamma_{pq}=& \sum_{\substack{r\\(r\neq p)}} |\bra{\phi_r} X \ket{\phi_p}|^2 G(\omega_{rp})+\sum_{\substack{r\\(r\neq q)}} |\bra{\phi_r} X \ket{\phi_q}|^2 G(\omega_{rq})\nonumber\\
&+|\bra{\phi_p} X \ket{\phi_p}+\bra{\phi_q} X \ket{\phi_q}|^2 G(0),
\end{aligned}
\right.
\end{equation}
where ${\cal P}$ denotes the Cauchy principal value.
Since $\Gamma_{pq}$ is positive, each off-diagonal element decays to be zero while oscillating at frequency, $\tilde{\epsilon}_p-\tilde{\epsilon}_q$;
\begin{equation}
\lim_{t\to \infty}\rho_{pq}(t)=0.
\end{equation}
Thus, the Floquet-Gibbs state is realized in the long-time asymptotic state (For the details, see~\cite{shirai2015condition}).

To reconsider the third condition we investigate systems with the Hamiltonian,
\begin{equation}
H_{\rm F}+H_{\rm B}+U^{\dagger}(t) X U(t) \otimes Y.
\end{equation}
This corresponds to the case where the first and second conditions are satisfied, but the third condition is not satisfied.
Here the third term describing the system-bath coupling is time periodic due to the second condition (see Eq.~(\ref{cond2})).
The Lindblad equation in this case reads for the diagonal elements,
\begin{equation}
\frac{d\rho_{pp}(t)}{dt}=\sum_{q} \sum_{n=-\infty}^{\infty} |\bra{\phi_p}X_n \ket{\phi_q}|^2 (G(\omega_{pqn})\rho_{qq}(t) - G(\omega_{qpn}) \rho_{pp}(t)),
\end{equation}
where $\omega_{pqn}=(\epsilon_p -\epsilon_q)/\hbar +n \Omega$ and
\begin{equation}
X_n =\frac{1}{T}\int_0^T U^{\dagger}(t) X U(t) e^{-i n \Omega t} dt.
\end{equation}
The difference from the case where the third condition is satisfied is found in the sum of $n$ in the transition probability, i.e.
\begin{equation}
T_{p \to q}= \sum_{n=-\infty}^{\infty} |\bra{\phi_p} X_n \ket{\phi_q}|^2 G(\omega_{qpn}).
\end{equation}
which generally breaks the detailed balance condition.
This breakdown of the Floquet-Gibbs state can be qualitatively understood as follows; The time dependence of the interaction Hamiltonian, $U^{\dagger}(t)H_{\rm SB}U(t)$, stimulates excitations inside the thermal bath around the frequency of $\Omega, 2\Omega,$ and so on.

However even when the third condition is not satisfied, it is expected that the Floquet-Gibbs state will appear as far as the response of the thermal bath to the high frequency field is weak.
In other words the correlation function of the thermal bath $G(\omega)$ decays as
\begin{equation}
G(\omega) \sim e^{-\frac{|\omega|}{\omega_c}},
\end{equation}
and $\omega_c \ll \Omega$.
In this case the transition probability is approximately written as
\begin{equation}
T_{p \to q} \simeq |\bra{\phi_p} X_0 \ket{\phi_q}|^2 G(\omega_{qp}).
\end{equation}
It has the same form as that when the third condition is satisfied (see Eq.~(\ref{transition})), and thus the detailed balance condition is recovered.
Therefore the third condition can be replaced by the following condition,
\begin{enumerate}[label={\arabic*}$^\prime.$]
\setcounter{enumi}{2}
\item The frequency of the driving field is much larger than the characteristic energy scale of the system-bath coupling, $\Omega \gg \omega_c$.
\end{enumerate}

\section{Effective Floquet-Gibbs state for systems with finite system-bath coupling}
In this section we discuss the relation between the dissipation effects whose strength is controlled by $\gamma$ (see Eq.~(\ref{relaxation_dynamics})) and the first condition $\hbar\Omega \gg \| H_0 \|$.
We here investigate the case where the second and the third conditions are satisfied, but the first condition is not satisfied.
The total Hamiltonian in the rotating frame is then given by
\begin{equation}
H_{\rm T, R}(t)=H_{\rm R}(t)+H_{\rm B}+H_{\rm SB},
\end{equation}
where $H_{\rm R}(t)$ is time periodic due to the second condition.

We first provide our expectation that the finite dissipation effects can lift the first condition, $\hbar\Omega \gg \| H_0 \|$.
For the realization of the Floquet-Gibbs state in a system with an infinitesimal system-bath coupling, this condition is necessary to avoid the energy absorption due to a resonance effect.
When this condition is broken there may exist two eigenstates in the spectral of $H_0$ which are in resonance with the driving field.
The system is then heating up and the long-asymptotic state is deviated from the Floquet-Gibbs state.
However when the heating effect is suppressed by the dissipation effect, it is expected that the Gibbs form will appear.

It is also known~\cite{mori2016rigorous} that the timescale for heating is extremely long when
\begin{equation}
\hbar \Omega \gg \text{( single site energy )},
\end{equation}
where the single site energy is of the order of the energy cost for local configuration changes, e.g. an excitation energy for flipping a spin on a site.
The energy is independent of the size of the system, which is denoted by $V$, and thus this condition is much weaker than the first condition, i.e. $\hbar \Omega \gg \| H_0 \| \sim O(V)$.
In the following we focus on this regime.

It is naively expected that the Floquet Hamiltonian $H_{\rm F}$ plays a role of the Hamiltonian in the expression of the Gibbs form also in this case, but there is a problem using it when the first condition is broken.
This is because the Floquet Hamiltonian is non-local, and its eigenstates are identical to the infinite temperature state~\cite{lazarides2014equilibrium,dalessio2014long-time,ponte2015periodically}.
Namely for local operators $O$ and each eigenstate of the Floquet Hamiltonian $\ket{\phi_p}$,
\begin{equation}
\bra{\phi_p}O\ket{\phi_p} \approx {\rm Tr} (O \rho_{\beta=0}^{\rm can}),
\end{equation}
where $\rho_{\beta=0}^{\rm can}$ is the infinite temperature state,
which is nothing but a totally random state.
However when the system is in contact with the thermal bath, the finite $\gamma$ suppresses the heating to infinite temperature.
Thus the Floquet Hamiltonian and its eigenbasis is inappropriate to describe the long-time asymptotic state.

To seek an alternative to the Floquet Hamiltonian, we employ the Floquet-Magnus expansion~\cite{blanes2009magnus}, which is $\Omega^{-1}$-expansion of the Floquet Hamiltonian.
We then obtain the truncated Floquet Hamitonian from this expansion up to the $n$th order,
\begin{equation}
H_{\rm F}^{(n)}=\sum_{k=0}^n \Omega_k,
\end{equation}
where $\| \Omega_k \|$ is the order of $\Omega^{-k}$.
The leading order term and the next-leading order term are explicitly given by
\begin{equation}
\left\{
\begin{aligned}
\Omega_0=&\frac{1}{T}\int_0^T H_{\rm R}(t) dt,\\
\Omega_1=&-\frac{i}{2T}\int_0^T dt_1 \int_0^{t_1} dt_2 [H_{\rm R}(t_1), H_{\rm R}(t_2)],\label{Magnus_eq}
\end{aligned}
\right.
\end{equation}
in which the Hamiltonian in the rotating frame $H_{\rm R}(t)$ is used.
The leading order term $H_{\rm F}^{(0)}$ is simply the time average of the rotating Hamiltonian, which approximately describes the isolated dynamics as discussed previously (see Eq.~(\ref{time_average})).
Recent extensive studies on the isolated periodically driven systems have shown that the appropriately truncated Floquet Hamiltonian describes long-lived transient states before reaching the infinite temperature state~\cite{kuwahara2016floquet,mori2016rigorous,abanin2017rigorous}.
In this sense it is reasonable to use the eigenbasis of the truncated Floquet Hamiltonian to probe the idea of the Gibbs form for finite dissipative systems.

\begin{figure}[t]
\begin{center}
\resizebox{0.3\columnwidth}{!}{
\includegraphics{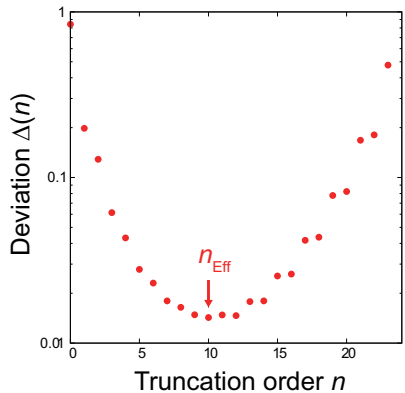} }
\end{center}
\caption{
(Color online) Truncation order $n$ in the Floquet-Magnus expansion vs Deviation between $H_{\rm F}^{(n_{\rm Eff})}$ and $H(t)$, $\Delta (n)$.
The minimum of $\Delta (n)$ determines the value of $n_{\rm Eff}$.
}
\label{Magnus}
\end{figure}

The appropriate truncation order is determined by minimizing the deviation between the time evolution of $H_{\rm F}^{(n)}$ and that of $H(t)$~\cite{kuwahara2016floquet}.
The deviation is quantitatively measured by
\begin{equation}
\Delta(n)=\| e^{-i H_{\rm F}^{(n)}T}-{\cal T} e^{-i\int_0^T H(t) dt} \|.
\end{equation}
In the ``high frequency'' regime where the frequency is much larger than the single site energy but the first condition, $\hbar \Omega \gg \| H_0 \|$, is broken, it typically shows the curve like Fig.~\ref{Magnus}.
The deviation initially decreases as the increase of the truncation order $n$, but after reaching the minimum it increases again.
The minimum determines the value of $n_{\rm Eff}$, and the truncated Floquet Hamiltonian approximately describes the driven dynamics,
\begin{equation}
H_{\rm T, R}(t) = H_{\rm R}(t)+H_{\rm B}+H_{\rm SB} \simeq H_{\rm F}^{(n_{\rm Eff})} +H_{\rm B}+H_{\rm SB}.
\end{equation}
We call here $H_{\rm F}^{(n_{\rm Eff})}$ effective Floquet Hamiltonian.

Since the total Hamiltonian is time independent, the system in contact with a thermal bath will be relaxed to the equilibrium state,
\begin{equation}
\rho_{\rm eq}=\frac{{\rm Tr_B}e^{-\beta (H_{\rm F}^{(n_{\rm Eff})} +H_{\rm B}+H_{\rm SB})}}{{\rm Tr}e^{-\beta (H_{\rm F}^{(n_{\rm Eff})} +H_{\rm B}+H_{\rm SB})}},\label{equilibrium}
\end{equation}
where ${\rm Tr_B}$ is the trace over the bath Hilbert space.
This expression is different from the Gibbs form of the bare Hamiltonian $H_{\rm F}^{(n_{\rm Eff})}$ due to the finite system-bath coupling.
In the framework of the quantum master equation,
we use the Redfield equation in order to take the weak but finite system-bath coupling into account.
The stationary solution of the Redfield equation is obtained in power series of $\gamma$ as
\begin{equation}
\rho_{\rm st}=\rho^{(0)} +\gamma \rho^{(1)} + \cdots.
\end{equation}
The leading order $\rho^{(0)}$ is obtained by
\begin{equation}
\left\{
\begin{aligned}
&\sum_{q} |\bra{\phi_p^{\rm E}}X \ket{\phi_q^{\rm E}}|^2 (G(\omega_{pq}^{\rm E})\rho_{qq}^{(0)} - G(\omega_{qp}^{\rm E}) \rho_{pp}^{(0)})=0,\\
&\rho_{pq}^{(0)} =0 \text{ for } p\neq q,
\end{aligned}
\right.
\end{equation}
where $\rho_{pq}^{(0)}=\bra{\phi_p^{\rm E}} \rho^{(0)} \ket{\phi_q^{\rm E}}$ is a matrix element of $\rho^{(0)}$ in the eigenbasis of $H_{\rm F}^{(n_{\rm Eff})}$, i.e. $H_{\rm F}^{(n_{\rm Eff})} \ket{\phi_p^{\rm E}}=\epsilon_p^{\rm E} \ket{\phi_p^{\rm E}}$, and $\omega_{pq}^{\rm E} = (\epsilon_p^{\rm E}-\epsilon_q^{\rm E})/\hbar$.
This form is equivalent to that of the Lindblad equation, and thus the solution is given by the Gibbs state of $H_{\rm F}^{(n_{\rm Eff})}$ (see the arguments from Eq.~(\ref{Lindblad}) to Eq.~(\ref{Boltzmann})).
The next leading order $\rho^{(1)}$ partially reproduces the equilibrium state $\rho_{\rm eq}$, and thus $\rho_{\rm st}$ is correct up to only the leading order of $\gamma$~\cite{mori2008dynamics}.
We here adopt from those satisfying $\rho = \rho_{\rm eq} +O(\gamma)$ a simple expression for the Effective Floquet-Gibbs state:
\begin{equation}
\rho_{\rm EFG} \equiv \frac{e^{-\beta H_{\rm F}^{(n_{\rm Eff})}}}{{\rm Tr} e^{-\beta H_{\rm F}^{(n_{\rm Eff})}}},
\end{equation}
which gives a good approximation of $\rho_{\rm eq}$ as far as the system-bath coupling is weak.
It is noted that the effective Floquet Hamiltonian is local, and thus it is totally different from the Floquet Hamiltonian $H_{\rm F}$ when the first condition is broken.

We here give a remark on the ``weak'' system-bath coupling.
When the Effective Floquet-Gibbs state appears, the dissipation effect overcomes the heating effect, and thus the relaxation timescale $(\propto \gamma^{-1})$ is no more the slowest timescale.
The weak-coupling limit, i.e. $\gamma \to 0$, is thus inappropriate to treat this situation.
In order to treat the comparable timescales within the master equation formalism, we have to use the Redfield equation instead of the Lindblad equation (see details in~\cite{shirai2016effective}).

Finally we demonstrate it in a spin-chain model.
The Hamiltonian reads
\begin{align}
&H(t)=H_0+\Omega H_{\rm ex}(\Omega t),\\
&\left\{
\begin{aligned}
H_0&=\sum_{i=1}^6 (h^z \sigma^z +h^x \sigma^x) -\sum_{i=1}^5 J \sigma_i^x \sigma_{i+1}^x,\\
\Omega H_{\rm ex}(\Omega t)&=\sum_{i=1}^6 \frac{\hbar \Omega}{3} \cos (\Omega t) \sigma_i^x,
\end{aligned}
\right.
\end{align}
where the strength of the exchange coupling is denoted by $J$.
On each spin in addition to static magnetic fields along $x$-axis and $z$-axis with the strength denoted by $h^x$ and $h^z$, respectively, an oscillating magnetic field along $x$-axis is applied.
It is noted that the second condition, i.e. $[H_{\rm ex}(t_1), H_{\rm ex}(t_2)]=0$, is satisfied.
In the following we ideally assume the third condition $[H_{\rm SB}, H_{\rm ex}(t)]=0$,
which is realized when the system is coupled to a thermal bath through $\sigma^x$.
Here we set the parameters as $(h^z, h^x, J)=(0.5, 0.35, 0.375)$.
We study the high frequency regime,
in which the frequency $\hbar \Omega =4.6$ is much larger than the single site energy, e.g. Zeeman energy $2 h^z=1$.
The value of the frequency is chosen so that there are two eigenstates of $H_{\rm F}^{(n_{\rm Eff})}$ in resonance with the driving field.
It is noted that even in the ``high frequency'' regime the first condition is not satisfied,
which is necessary to study the competition between the heating effect and the dissipation effect.
Namely the frequency is set to be smaller than the spectral width of $H_0 (=7.6)$.

In order to probe the difference between the long-time asymptotic state $\rho_{\rm asy}$ (see Eq.~(\ref{asymptotic})) and the Effective Floquet-Gibbs state $\rho_{\rm EFG}$, we calculate the trace distance,
\begin{equation}
\Delta {\rm Prob} ={\rm Tr} |\rho_{\rm asy}-\rho_{\rm EFG}|.
\end{equation}
This gives a bound on the difference between the expectation values of an observable $O$ over the two distributions, $\rho_{\rm asy}$ and $\rho_{\rm EFG}$;
\begin{equation}
\braket{O}_{\rm asy} -\braket{O}_{\rm EFG} \equiv {\rm Tr (O \rho_{\rm asy})}-{\rm Tr (O \rho_{\rm EFG})} \leq \|O \| \Delta {\rm Prob}.
\end{equation}
Thus the smallness of the difference in $\Delta {\rm Prob}$ ensures that a measurement gives a value close to that for the Effective Floquet-Gibbs state.

\begin{figure}[t]
\begin{center}
\resizebox{0.3\columnwidth}{!}{
\includegraphics{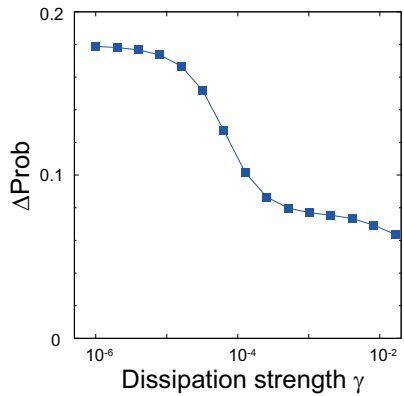} }
\end{center}
\caption{
(Color online) Deviation of the long-time asymptotic state and the Effective Floquet-Gibbs state, $\Delta {\rm Prob}$ vs Dissipation strength $\gamma$.
}
\label{EFG}
\end{figure}

Finally we show in Fig.~\ref{EFG} the deviation $\Delta {\rm Prob}$ vs the dissipation strength $\gamma$.
For small $\gamma$ large deviation appears due to the resonance effect,
but as $\gamma$ is increased the deviation is decreased.
This clearly describes the dissipation effect which pushes the long-time asymptotic state into the Effective Floquet-Gibbs state. 

\section{Conclusion}
In this paper we discussed the long-time asymptotic states of periodically driven open quantum systems.
We first investigated a system with an infinitesimal system-bath coupling, i.e. $\gamma \to 0$,
and showed that the under the three conditions labeled by 1, 2, and 3, the Floquet Hamiltonian $H_{\rm F}$ is thermodynamically relevant, i.e. $\rho_{\rm asy} \simeq e^{-\beta H_{\rm F}}/{\rm Tr} e^{-\beta H_{\rm F}}$.
We also formulated a Lindblad type of quantum master equation in a rotating frame, and showed that the condition 3, $[H_{\rm SB}, H_{\rm ex}(t)]=0$, can be lifted by taking a timescale of the thermal bath into account.
We next discussed the $\gamma$ dependence of the long-time asymptotic state.
We provided our expectation that with an aid of the dissipation effect, the effective Floquet Hamiltonian $H_{\rm F}^{(n_{Eff})}$ defined by the Floquet-Magnus expansion is thermodynamically relevant when $\hbar \Omega \gg \text{(single site energy)}\sim O(V^0)$ instead of the condition 1, $\hbar \Omega \gg \| H_0 \| \sim O(V)$.
We showed that this expectation is supported in a numerical simulation of a spin chain model. 

Here we reconsidered the theory of the Floquet-Gibbs state~\cite{shirai2015condition,shirai2016effective} in the rotating frame.
This perspective will be helpful for the further investigation on the competitive phenomena due to the excitations by periodically driving fields and the dissipation effect. 




\end{document}